\documentclass[aps,pre,twocolumn,superscriptaddress,showpacs,floatfix]{revtex4-1}
\usepackage[utf8]{inputenc}

\usepackage{graphicx}  
\usepackage{dcolumn}   
\usepackage{bm}        
\usepackage{amssymb,times}

\usepackage{amsmath}
\usepackage{amsthm}

\usepackage{graphicx}

\begin{document}

\title{Layer degradation triggers an abrupt structural transition in multiplex networks}

\author{Emanuele Cozzo}
\affiliation{Institute for Biocomputation and Physics of Complex Systems (BIFI), University of Zaragoza, Zaragoza 50009, Spain}
\affiliation{Department of Theoretical Physics, University of Zaragoza, Zaragoza 50009, Spain}

\author{Guilherme Ferraz de Arruda}
\affiliation{ISI Foundation, Via Chisola 5, 10126 Torino, Italy}

\author{Francisco A. Rodrigues}
\affiliation{Departamento de Matem\'{a}tica Aplicada e Estat\'{i}stica, Instituto de Ci\^{e}ncias Matem\'{a}ticas e de Computa\c{c}\~{a}o, Universidade de S\~{a}o Paulo - Campus de S\~{a}o Carlos, Caixa Postal 668, 13560-970 S\~{a}o Carlos, SP, Brazil.}
\affiliation{Mathematics Institute, University of Warwick, Gibbet Hill Road, Coventry CV4 7AL, UK.}
\affiliation{Centre for Complexity Science, University of Warwick, Coventry CV4 7AL, UK.}

\author{Yamir Moreno}
\affiliation{Institute for Biocomputation and Physics of Complex Systems (BIFI), University of Zaragoza, Zaragoza 50009, Spain}
\affiliation{Department of Theoretical Physics, University of Zaragoza, Zaragoza 50009, Spain}
\affiliation{ISI Foundation, Via Chisola 5, 10126 Torino, Italy}

\date{\today}

\begin{abstract}
Network robustness is a central point in network science, both from a theoretical and a practical point of view.
In this paper, we show that layer degradation, understood as the continuous or discrete loss of links' weight, triggers a structural transition revealed by an abrupt change in the algebraic connectivity of the graph. Unlike traditional single layer networks, multiplex networks exist in two phases, one in which the system is protected from link failures in some of its layers and one in which all the system senses the failure happening in one single layer. We also give the exact critical value of the weight of the intra-layer links at which the transition occurs for continuous layer degradation and its relation with the value of the coupling between layers. This relation allows us to reveal the connection between the transition observed under layer degradation and the one observed under the variation of the coupling between layers.
\end{abstract}

\maketitle

\section{Introduction}

Multilayer networks have gained a lot of attention in the last years \cite{ReviewKivela,CozzoBook}. Within the class of multilayer networks, multiplex networks are those that are made up of a set of nodes that can interact by means of different types of interactions, each of which forms a network, called layer. The coupling between different layers accounts for the fact that some -- or all -- of the nodes participate in more than one layer \cite{ReviewKivela,CozzoBook}. Many real-world complex systems are better described in terms of multiplex networks rather than of aggregated (single-layer) complex networks  \cite{barret2012taking,cozzo2015structure,cardillo2013modeling}. Besides, much of the theoretical interest on these networks arises because multiplex networks display very different properties with respect to traditional (single-layer) networks. In this regard, much attention has been devoted to the study of the impact of the coupling on the structural organization of a multiplex network \cite{Cozzo2016Characterization,Sahneh2015,Estrada2014}.

Network robustness -- intended in a general way -- has been a major topic in network science \cite{callaway2000robustness,scott2006robustness,estrada2006robustness,memmott2004tolerance}.
In general, from a theoretical perspective, it deals with the study of how network invariants change under structural perturbations, such as the removal or addition of nodes and/or links, or the modification of the strengths of the links. In particular, it is of interest to study the robustness of a network under degradation, by which we mean: (i) random links removal (failures), (ii) deterministic links removal (attacks) or (iii) the lowering of links' strength (continuous degradation). These network degradations describe real-world processes like failures in traffic networks \cite{wu2007failures} or neurodegenerative diseases \cite{Selkoe2002alzheimer}, among many others. 

The algebraic connectivity (i.e., the second-smallest eigenvalue of the Laplacian of a connected graph)~\cite{Fiedler} is a good measure of network robustness since it measures the extent to which it is difficult to cut the network into independent components \cite{Jamakovic2008}. In fact, it is a lower bound for both the edge connectivity and node connectivity of a graph, that is, respectively the minimal number of edges and nodes that have to be removed to disconnect the graph. Moreover, it is also a good measure from a dynamical point of view. For example, the time needed to synchronize a network of oscillators is also related to the algebraic connectivity \cite{Almendral2007dynamical}, as well as the time scale of diffusion processes \cite{gomez2013diffusion,ATejedor2018}. Thus, in this sense, the algebraic connectivity represents the connection between the structural and the dynamical robustness of a network.

The algebraic connectivity of a multiplex network is defined as the first non-zero eigenvalue of the supra-Laplacian \cite{ReviewKivela,CozzoBook}
\begin{equation}
\bar{\mathcal{L}}=\bigoplus_\alpha \mathbf{L}^{(\alpha)} + p \mathcal{L}_C,
\end{equation}
where $\mathbf{L}^{(\alpha)}$ is the Laplacian of the network in layer $\alpha$, $\mathcal{L}_C$ is the Laplacian of the coupling graph, i.e. the graph formed by the edges between the same nodes in different layers, and $p$ is the coupling weight.
If we label the eigenvalues $\{\bar{\mu}_i\}$ of $\bar{\mathcal{L}}$ in ascending order, $\bar{\mu}_2$ is the algebraic connectivity of the given multiplex network. Its dependency on the value of the coupling weight $p$ has been studied \cite{DAgostino,RadicchiArenas}, and it was found that $\bar{\mu}_2$ follows two distinct regimes when varying the coupling parameter $p$:
\begin{equation}
\bar{\mu}_2 =
        \begin{cases}
            mp, &\ if\ p\leq p^* \\
            \leq \mu_2(\mathbf{L}_a), &\ if\ p>p^*
        \end{cases},
\end{equation}
where $m$ is the number of layers, and $\mu_2(\mathbf{L}_a)$ is the algebraic connectivity of the aggregate network \cite{Sanchez2014}, the critical value $p^*$ is given by the crossing of two eigenvalues of $\bar{\mathcal{L}}$ \cite{Cozzo2016Characterization}. In other words, the critical point $p^*$ can be defined as a point of non-analyticity of the algebraic connectivity as a function of the coupling parameter $p$ \cite{DAgostino} and this particular behavior affects both the structural organization of a multiplex network \cite{RadicchiArenas,Cozzo2016Characterization} and the dynamical behavior of processes defined on it \cite{Sahneh2015,gomez2013diffusion,ATejedor2018}. Thus, for a node-aligned multiplex network, when varying the coupling parameter $p$, before $p^*$ the minimum cut only implies inter-layer couplings, while after $p^*$ intra-layer links are implied. In one case, the cut is ``parallel'' to the layers, in the others, it traverses them.

In this work, we show that the algebraic connectivity of a multiplex network also follows two distinct regimes during the process of layer degradation. The system experiments an abrupt structural transition as in the case of the transition experimented when varying the coupling parameter $p$, due to the crossing of two eigenvalues. More interestingly, unlike the structural transition experimented when the coupling parameter varies, during the layer degradation it remains constant for a finite fraction of links removed as well as for a finite interval of variation of the intra-layer weights before it starts to decrease. This also differentiates the behavior of a multiplex network from that of a traditional (single-layer) complex network.

\section{Continuous layers degradation}

In this section, we study the continuous layer degradation. In order to model this process, we introduce a set of intra-layer weight parameters, $\{t_\alpha\}$, where $t_\alpha$ are the intra-layer weights in layer $\alpha$. Moreover, we fix the coupling parameter $p$ in the disconnected phase, formally, $p$ is fixed at a given $p_0<p^*$. Consequently, the supra-Laplacian now reads
\begin{equation}
\bar{\mathcal{L}}=\bigoplus_\alpha t_\alpha \mathbf{L}^{(\alpha)} + p_0\mathcal{L}_C 
\label{weightedsupraL}
\end{equation}
and the algebraic connectivity is $\bar{\mu}_2=mp_0$. Without loss of generality, we set all the $t_\alpha$s equal to the unity, but one, which is set as $t_\delta=t$. In particular, we chose the layer $\delta$ as the layer with the lowest individual algebraic connectivity $\mu_2^{(\delta)}$. We call it the Laplacian dominant layer (in line with the language used in \cite{Cozzo2013} for the case of the supra-Adjacency matrix). By construction, the algebraic connectivity for $t=1$ is $\bar{\mu}_2=mp_0$, while the next eigenvalue can be approximated as \cite{Milanese2010}
\begin{equation}
\bar{\mu}_3\sim t\mu_2^{(\delta)}+(m-1)p_0.
\label{mu3approx}
\end{equation}
The eigenvalue $\bar{\mu}_3$ decreases with $t$ and, for continuity, at a given point $t^*$, it hits the value $mp_0$. Thus, we can conclude that the algebraic connectivity follows two distinct regimes:
\begin{equation}
\bar{\mu}_2=\begin{cases}
        mp_0, &\ if\ t\geq t^*\\
        \sim t\mu_2^{(\delta)}+(m-1)p_0, &\ if\ t<t^*\\
        \end{cases}.
\end{equation}
Actually, the r.h.s. of Eq. \eqref{mu3approx} is an upper bound for $\bar{\mu}_3$ \cite{Sidney2013} and by equating it to $mp_0$ we get a lower bound for the point $t^*$ at which the algebraic connectivity enters a distinct regime:
\begin{equation}
t^*>\frac{p_0}{\mu_2^{(\delta)}}.
\label{tstarbound}
\end{equation}

\begin{figure}[t]
\includegraphics[width=\columnwidth]{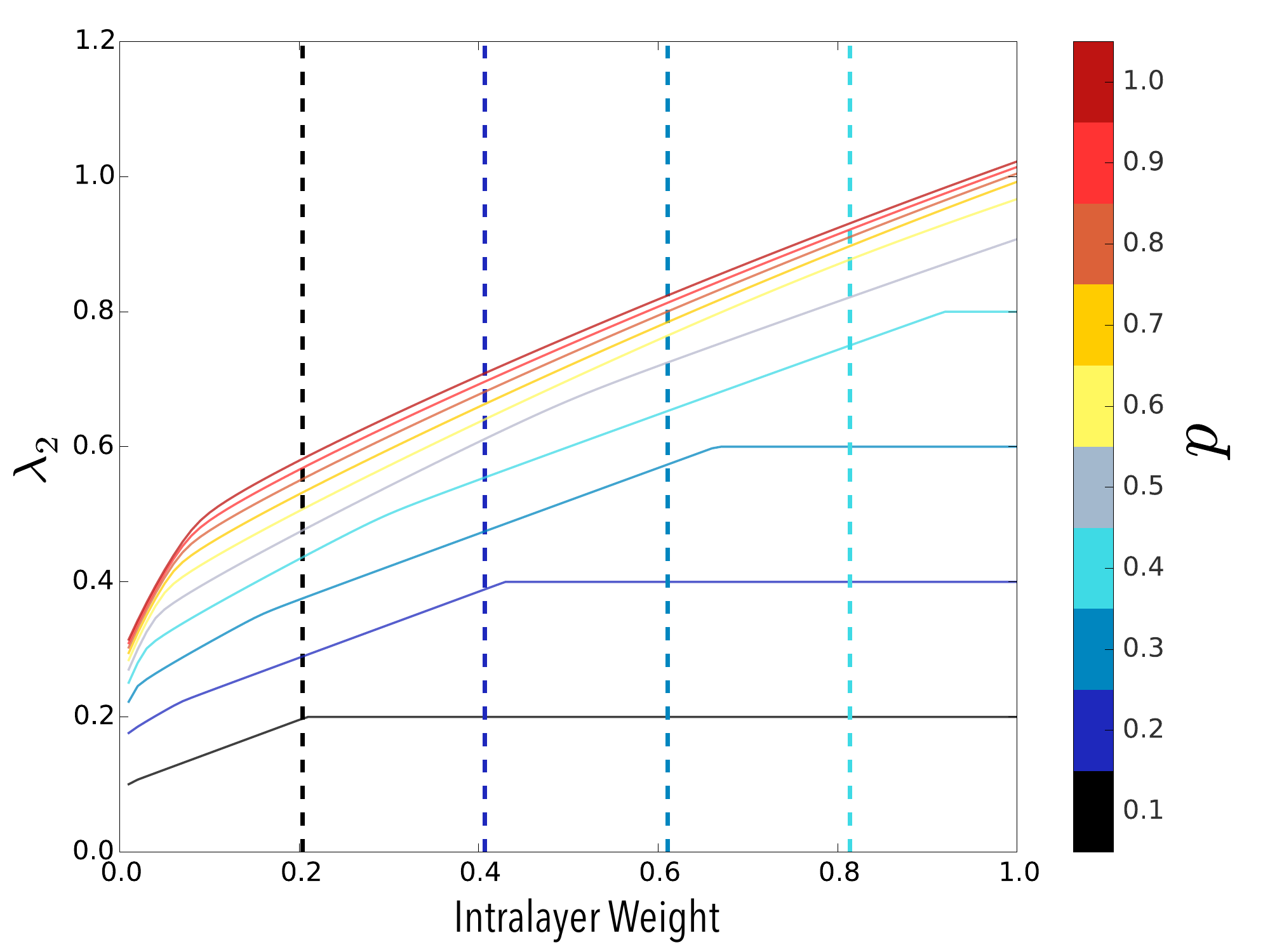}
\caption{Continuous degradation of the Laplacian dominant layer in a multiplex network composed by two Erd\"os -- Rényi networks with $N=100$ nodes and average degree $\langle k \rangle \approx 5$.}
\label{fig:continousdeg}
\end{figure}

As we can observe in Fig.\ref{fig:continousdeg}, the bound \eqref{tstarbound} is sharp for low values of $p_0$, where the approximation \eqref{mu3approx} is accurate. Besides, we can observe that the transition only exits when $p_0<p^*$, while for larger values, the algebraic connectivity is already in the regime in which it smoothly decreases (this happens for values of $p_0$ larger than $0.4$ in the particular settings of Fig. \ref{fig:continousdeg}). We can understand this behavior, as well as the mechanism that triggers the structural transition, by calculating the exact value of $t^*$ in the identical intra-layer weight case. This scenario is explored in the next section.

\subsection{Exact value of $t^*$ for identical weights}

Consider the case in which all the intra-layer weight parameters $t_\alpha$ are identical, i.e. $t_\alpha=t,\ \forall \alpha$. In this scenario, we can obtain the exact transition point, $t^*$, by reformulating the eigenvalue problem for the supra-Laplacian in terms of a polynomial eigenvalue problem \cite{Arruda2018poly}. In general, a polynomial eigenvalue problem is formulated as an equation of the form
\begin{equation}
\mathbf{Q}(\lambda)\mathbf{x}=0
\label{general}
\end{equation}
where $\mathbf{Q}(\lambda)$ is a polynomial matrix, whose elements are polynomials in $\lambda$ and its solutions are given by  
\begin{equation}
\det(\mathbf{Q}(\lambda))=0.
\label{generalsolution}
\end{equation}

The eigenvalue problem for the supra-Laplacian of a multiplex composed by two layers is expressed as
\begin{eqnarray}
\bar{\mathcal{L}}\mathbf{x}=
\left[
\begin{array}{c|c}
\mathbf{L}_a & -p\mathbf{I} \\
\hline
-p\mathbf{I} & \mathbf{L}_b
\end{array}
\right]
\left[
\begin{array}{c}
\mathbf{x}_a\\
\hline
\mathbf{x}_b
\end{array}
\right]
= \lambda
\left[
\begin{array}{c}
\mathbf{x}_a\\
\hline
\mathbf{x}_b
\end{array}
\right]
=\lambda\mathbf{x},
\end{eqnarray}
from which we get the following system of equations
\begin{eqnarray}
\mathbf{L}_a\mathbf{x}_a - p\mathbf{I}\mathbf{x}_b&=&\lambda\mathbf{x}_a\nonumber\\
\mathbf{L}_b\mathbf{x}_b - p\mathbf{I}\mathbf{x}_a&=&\lambda\mathbf{x}_b.
\label{quadratic}
\end{eqnarray}
Next, isolating $\mathbf{x}_a$ from the second equation we get
\begin{equation}
\mathbf{x}_a=-\frac{1}{p}(\mathbf{L}_b-\lambda{I})\mathbf{x}_b.
\end{equation}
Then, plugging the expression of $\mathbf{x}_a$ in the first equation of the system \eqref{quadratic}, we obtain
\begin{equation}
\left[\frac{1}{p}(\mathbf{L}_a-\lambda\mathbf{I})(\mathbf{L}_b-\lambda\mathbf{I})-p\mathbf{I}\right]\mathbf{x}_b=0,
\end{equation}
in which we can recognize a quadratic eigenvalue problem
\begin{equation}
\mathbf{Q}(\lambda)=\mathbf{A}\lambda^2+\mathbf{B}\lambda+\mathbf{C}=0
\end{equation} 
with
\begin{eqnarray}
&\mathbf{A}=& \mathbf{I} \nonumber\\
&\mathbf{B}=&-(\mathbf{L}_a+\mathbf{L}_b+2p\mathbf{I})\nonumber\\
&\mathbf{C}=&\mathbf{L}_a\mathbf{L}_b+p(\mathbf{L}_a+\mathbf{L}_b).
\end{eqnarray}
We know \cite{Sanchez2014} that $2p$ is always an eigenvalue of $\bar{\mathcal{L}}$ and therefore it is always also a solution of $\det(\mathbf{Q}(\lambda))=0$. Thus, we have
\begin{equation}
0 = \det(\mathbf{Q}(\lambda)) = \det(\mathbf{L}_a+\mathbf{L}_b)\det(\mathbf{L}_a\mathbf{L}_b(\mathbf{L}_a+\mathbf{L}_b)^\dagger-p\mathbf{I}).
\label{2psolution}
\end{equation}
The first term of the r.h.s is always $0$. Moreover, given that $2p$ is always a solution, the other term gives the crossing points for the eigenvalues, the first of which defines $p^*$. Thus, we have
\begin{equation}
\det(\mathbf{L}_a\mathbf{L}_b(\mathbf{L}_a+\mathbf{L}_b)^\dagger-p\mathbf{I}) = 0
\label{peigenvalue}
\end{equation}
that is, again, an eigenvalue problem in terms of $p$. Then, we have that the exact value of $p^*$ is given by
\begin{equation}
p^*=\lambda_2(\mathbf{L}_a\mathbf{L}_b(\mathbf{L}_a+\mathbf{L}_b)^\dagger).
\label{exactp}
\end{equation}

Now accounting for the weights $t_a = t_b = t$ and fixing the inter-layer coupling, $p=p_0$, Eq. \eqref{peigenvalue} reads
\begin{eqnarray}
\det(t_a\mathbf{L}_a t_b\mathbf{L}_b(t_a\mathbf{L}_a+t_b\mathbf{L}_b)^\dagger-p_0\mathbf{I}) &=& \nonumber \\
\det(t\mathbf{L}_a\mathbf{L}_b(\mathbf{L}_a+\mathbf{L}_b)^\dagger-p_0\mathbf{I}) &=& 0,
\end{eqnarray}
and consequently implying that
\begin{equation}
t^*=\frac{p_0}{\lambda_2( \mathbf{L}_a\mathbf{L}_b(\mathbf{L}_a+\mathbf{L}_b)^\dagger)}=\frac{p_0}{p^*}.
\label{exactt}
\end{equation}

In other words, we look for a $p_0$ that is the first non-zero eigenvalue of $H(t)=t\mathbf{L}_a\mathbf{L}_b(\mathbf{L}_a+\mathbf{L}_b)^\dagger$, i.e., we want the value $t^*$ such that
\begin{equation}
p_0=\lambda_2(H(t^*))= t^* = \lambda_2(\mathbf{L}_a\mathbf{L}_b(\mathbf{L}_a+\mathbf{L}_b)^\dagger),
\end{equation}
from which we get Eq. \eqref{exactt}.

\subsection{General mechanism}

\begin{figure}[t]
\includegraphics[width=\columnwidth]{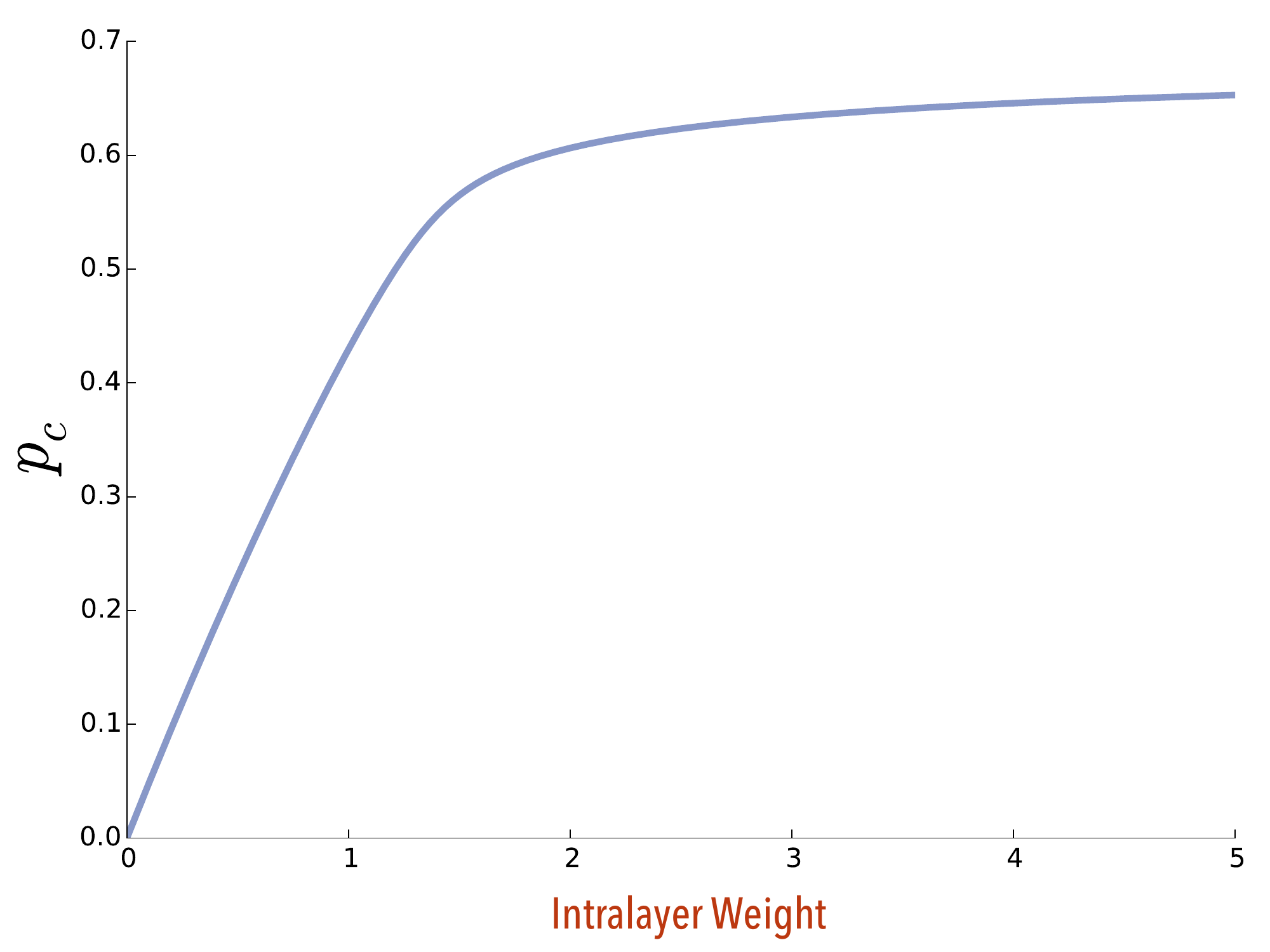}
\caption{Dependency of $p_c$ on the extreme value of the intra-layer weight for the same setting of Fig.\ref{fig:continousdeg}.}
\label{pc}
\end{figure}

Generally, it is always possible to write an equation of the form
\begin{equation}
p_0=\lambda_2(H(\{t^*_\alpha\}))
\label{generalt}
\end{equation}
that implicitly defines the structural transition point $\{t^*_\alpha\}$, which is the solution to the above parameter inverse eigenvalue problem. In other words, the triggering mechanism behind the structural transition, similar to the transition in $p$, is an eigenvalue crossing resulting from the fact that the actual value of the coupling $p$ is the first non-zero eigenvalue of the matrix $H(\{t\})$. Note, however, that the transition in $p$ can be directly obtained by simply varying $p$. Formally speaking, it can be found by solving a direct eigenvalue problem. On the other hand, for the transition in $t$, the transition point can be obtained by changing the values of the intra-layer weights until $p_0$ is the first non-zero eigenvalue of $H(\{t\})$ $-$ e.g., it is a parameter inverse eigenvalue problem \cite{Chu}.

The case of identical layers is the only one in which we can give an explicit equation for the transition point,  $t^*$, as in Eq. \eqref{exactt}. However, the parameter inverse eigenvalue problem can be solved numerically always. In our model, we consider that the weights are constrained in the interval $0< t_\alpha\leq 1$. This implies that there exists a value of the coupling parameter $p_0$ above which it is impossible to observe a transition in $t$. In fact, if we consider Eq. \eqref{exactt} for a value of $p_0>p^*$ we have
\begin{equation}
t^*=\frac{p_0}{p^*}>1.
\end{equation}
In general, given a range of variation for the intra-layer weights, there will always exist a value $p_c$ of the coupling parameter above which it is impossible to observe a transition in $t$. Thus, $p_c$ can be calculated by solving 
\begin{equation}
p_0=\lambda_2(H(\{\bar{t}_\alpha\})),
\end{equation}
where $\bar{t}_\alpha$ is the extremal values of $t_\alpha$ in its range of variation. For the setting of Fig.\ref{fig:continousdeg}, in which we have two layers, one of which with fixed intra-layer weight equal to $1$, we have that the dependency of $p_c$ on the extreme value of the intra-layer weight is as depicted in Fig. \ref{pc}.

\section{Links failure and attacks}

\begin{figure}[t]
\includegraphics[width=\columnwidth]{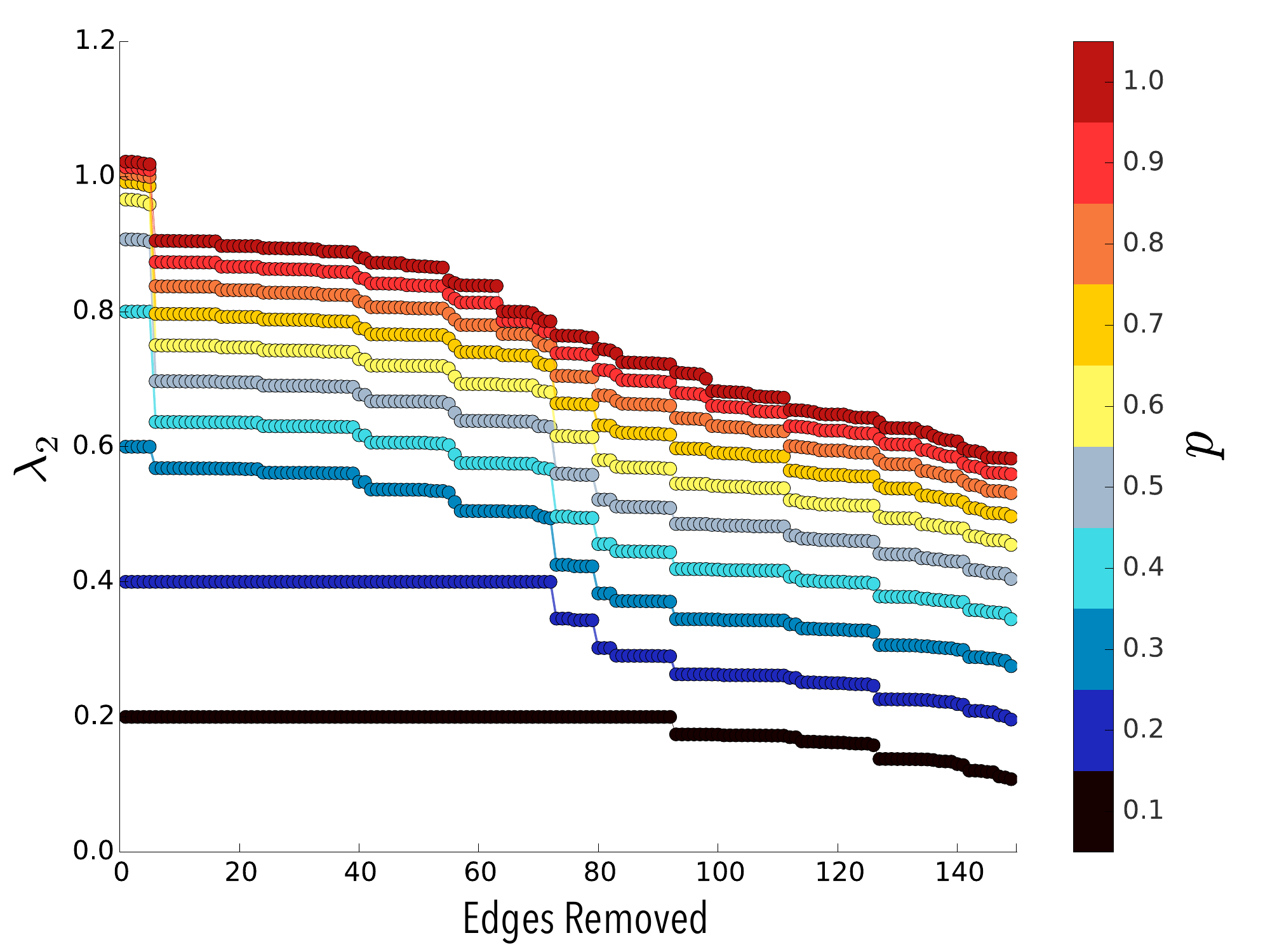}
\includegraphics[width=\columnwidth]{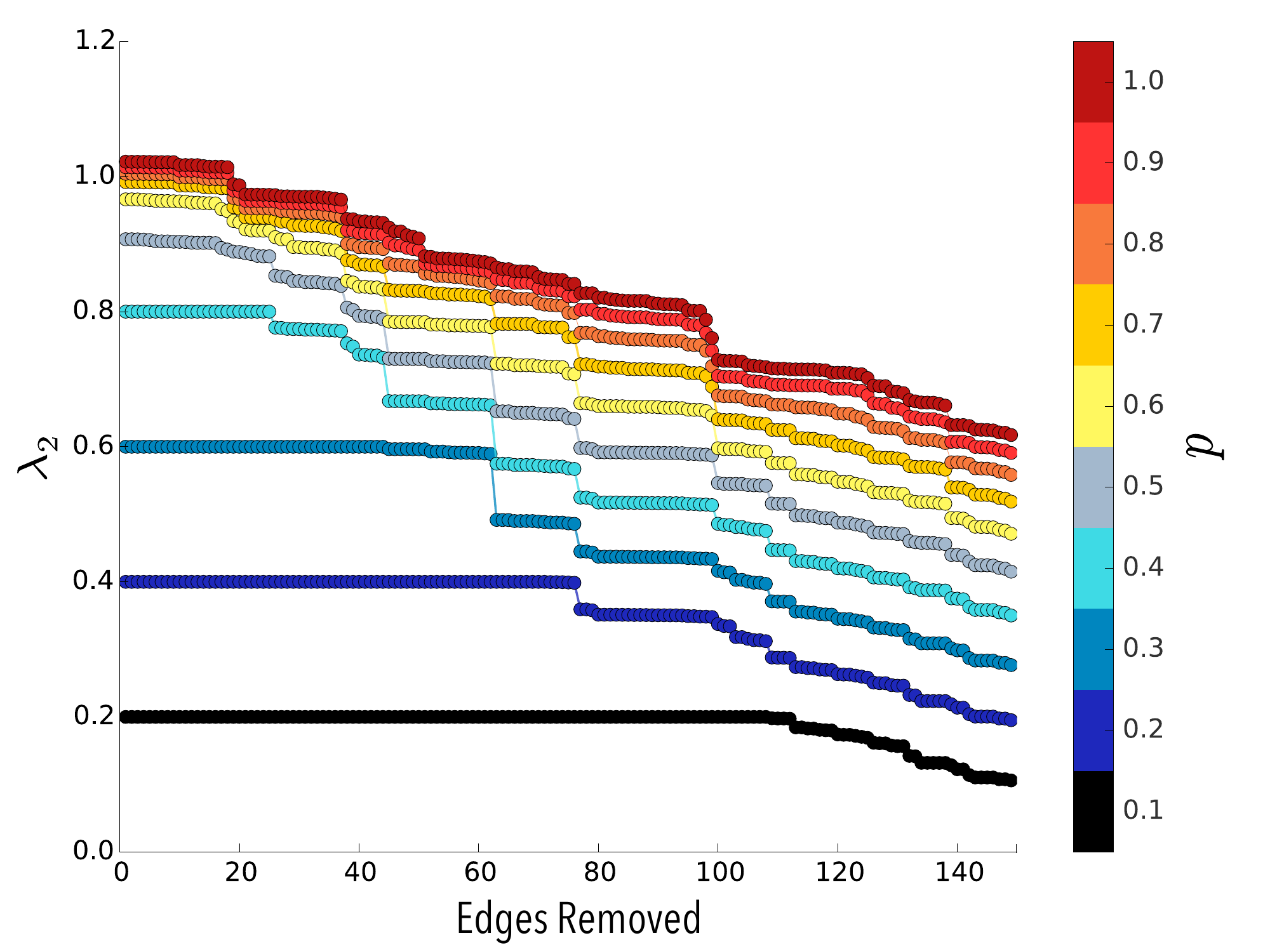}
\caption{Algebraic connectivity subject to attacks: the discrete layer degradation with descending (upper panel) and ascending (lower panel) ordering of the links according to $r$. The multiplex network used in this experiment is similar to the one used in Fig. \ref{fig:continousdeg}.}
\label{fig:attack}
\end{figure}

\begin{figure}[t]
\includegraphics[width=\columnwidth]{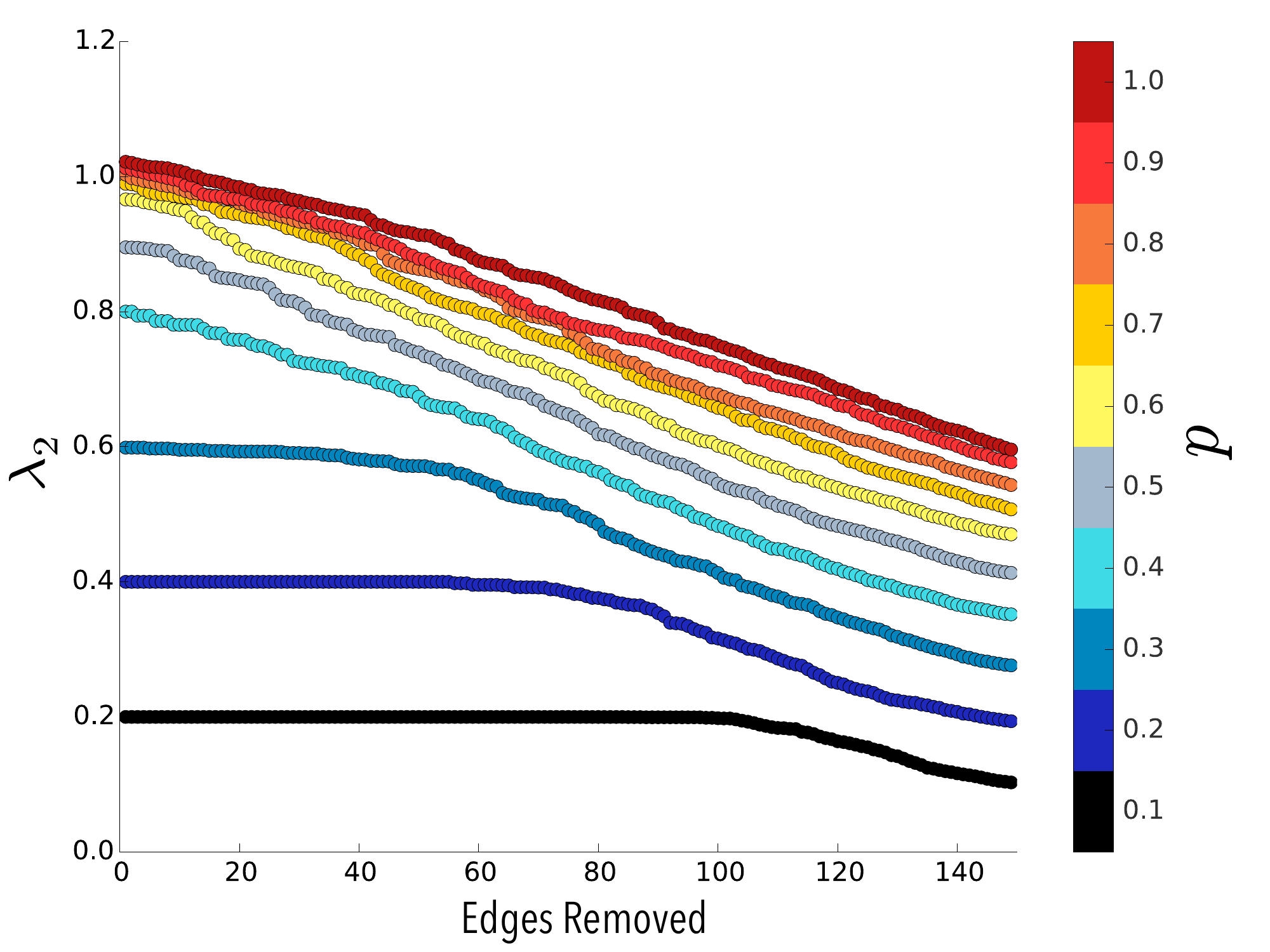}
\includegraphics[width=\columnwidth]{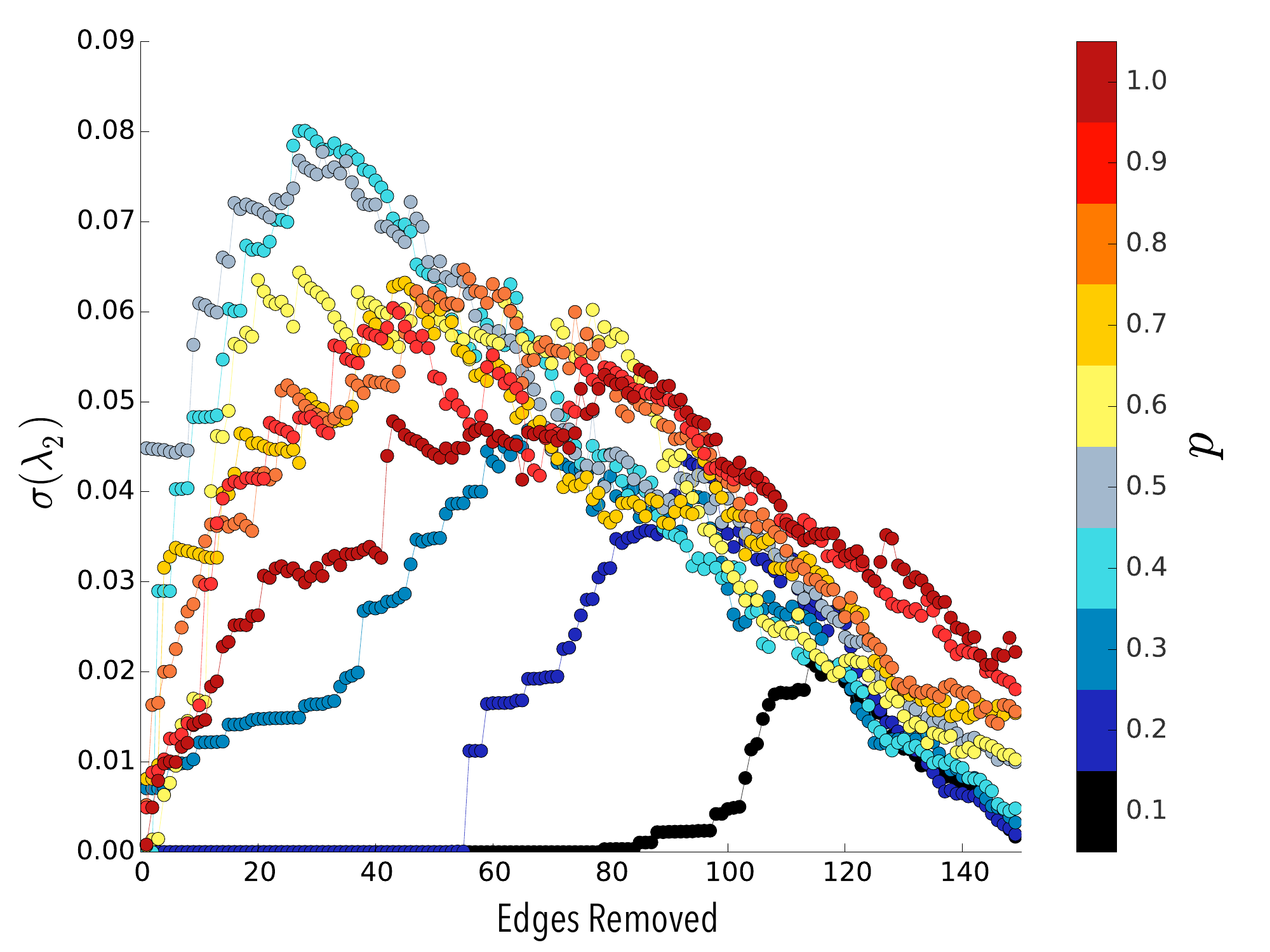}
\caption{Algebraic connectivity subject to random failures: discrete layer degradation with random removals of links (upper panel) and its respective standard deviation (lower panel). The multiplex network used in this experiment is similar to the one used in Fig. \ref{fig:continousdeg}.}
\label{fig:rndeig}
\end{figure}

In this section, we focus on the case of discrete layer degradation, which is given by the removal of links of the multiplex network. As in the previous section, we fix the layer in which we will remove links as the one with the lowest algebraic connectivity. In other words, the Laplacian dominant layer. Note that we take the layer $\delta$ as isolated and remove an arbitrary edge, $e$, which connects nodes $i$ and $j$. Thus, its algebraic connectivity can be approximated by
\begin{equation}
\mu_2^{(\delta-e)}\sim\mu_2^{(\delta)}-r^2,
\label{isolatedalgebraicapprox}
\end{equation}
where $\mu_2^{(\delta-e)}$ is the algebraic connectivity of the layer $\delta$ after removing link $e$ and $r=(x^{(\delta)}_{2i}-x^{(\delta)}_{2j})$, where $x_i$ is the $i-$th element of the Fiedler eigenvector, $\mathbf{x}^{(\delta)}_2$, which is associated with the algebraic connectivity, $\mu_2^{(\delta)}$.

As before, for a given $p_0<p^*$ the algebraic connectivity is $\bar{\mu}_2=mp_0$, whereas the next eigenvalue when a link $e$ is removed can be approximated as
\begin{equation}
\bar{\mu}_3\sim \mu_2^{(\delta-e)}+ (m-1)p_0.
\label{approxm3edges}
\end{equation}
In general, when a set $E$ of links is removed we have $r=\sum_{ij}(x^{(\delta)}_{2i}-x^{(\delta)}_{2j})$, where the sum is over all the links in $E$ \cite{Milanese2010}.
Since Eq. \eqref{approxm3edges} is an upper bound, as before, we obtain a lower bound for the critical value of $\mu_2^{(\delta-E)}$ from which the removal of an edge will cause a drop in the algebraic connectivity, i.e.,
\begin{equation}
\mu_2^{(\delta-E)*} > mp_0.
\end{equation}
Therefore, we have two different scenarios: (i) the targeted removal of edges, which we call attacks and (ii) the random removal of edges, which we call failures.

Firstly, let us analyze the attacks. Looking at Eq. \eqref{approxm3edges}, we have two evident strategies that are based on the entries of the eigenvector $x_2^{(\delta)}$. After ranking the links according to their associated value of $r$ we can remove them in ascending or descending order. We remark that we do not allow the network to break into more than one component. The critical fraction of edges that has to be removed in order to cause a drop in the algebraic connectivity is obviously larger in the second case, as can also be observed in Fig. \ref{fig:attack}. 

Secondly, regarding failures, links are randomly removed from the Laplacian-dominant layer. We also remark that during the random removal we do not allow that the network breaks into more than one component. We show these results in Fig. \ref{fig:rndeig}. In the lower panel of Fig. \ref{fig:rndeig} we observe the variation of the standard deviation of the algebraic connectivity. When $p_0$ is low enough in order to have a structural transition, as expected, the standard deviation is constantly zero until the critical fraction of removed edges is reached.

Additionally, it is also instructive to compare figures \ref{fig:attack} and \ref{fig:rndeig}. Obviously, the attacks in ascending order produce a much more abrupt change if compared to the descending order. Furthermore, on the failure case, we have a much smoother curve as a consequence of taking averages. Note that the final result for the three methods is nonetheless similar. This is because, as said above, we do not allow that the network breaks into more than one component. Thus, the final network should be approximately a minimum spanning tree of the original degraded network.

\section{The Shannon entropy of the Fiedler vector}

\begin{figure}[t]
\includegraphics[width=\columnwidth]{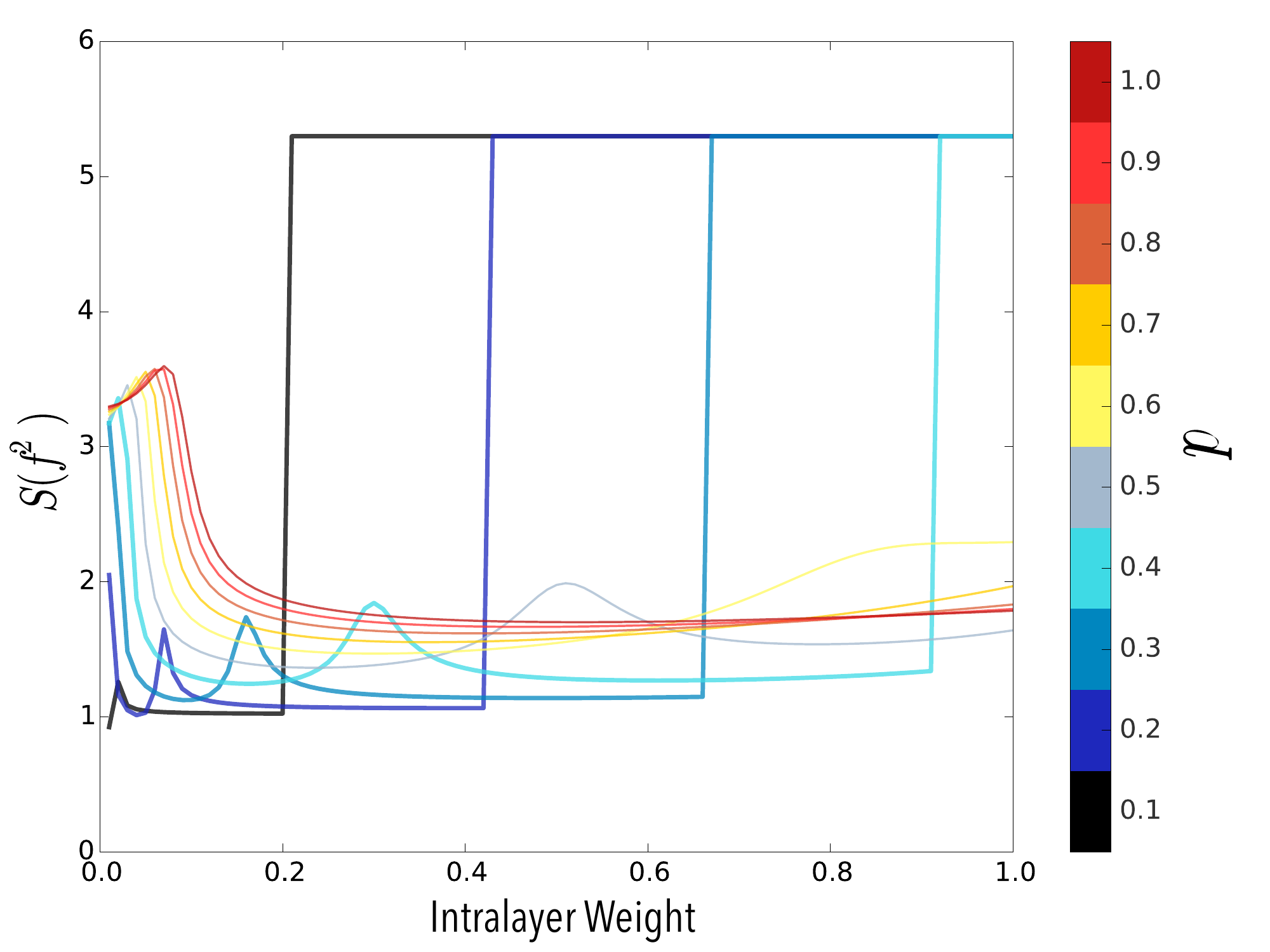}
\caption{Shannon entropy of the Fiedler vector under continuous layer degradation. The multiplex network used in this experiment is similar to the one used in Fig. \ref{fig:continousdeg}.}
\label{fig:continousdegentr}
\end{figure}

\begin{figure}[t]
\includegraphics[width=\columnwidth]{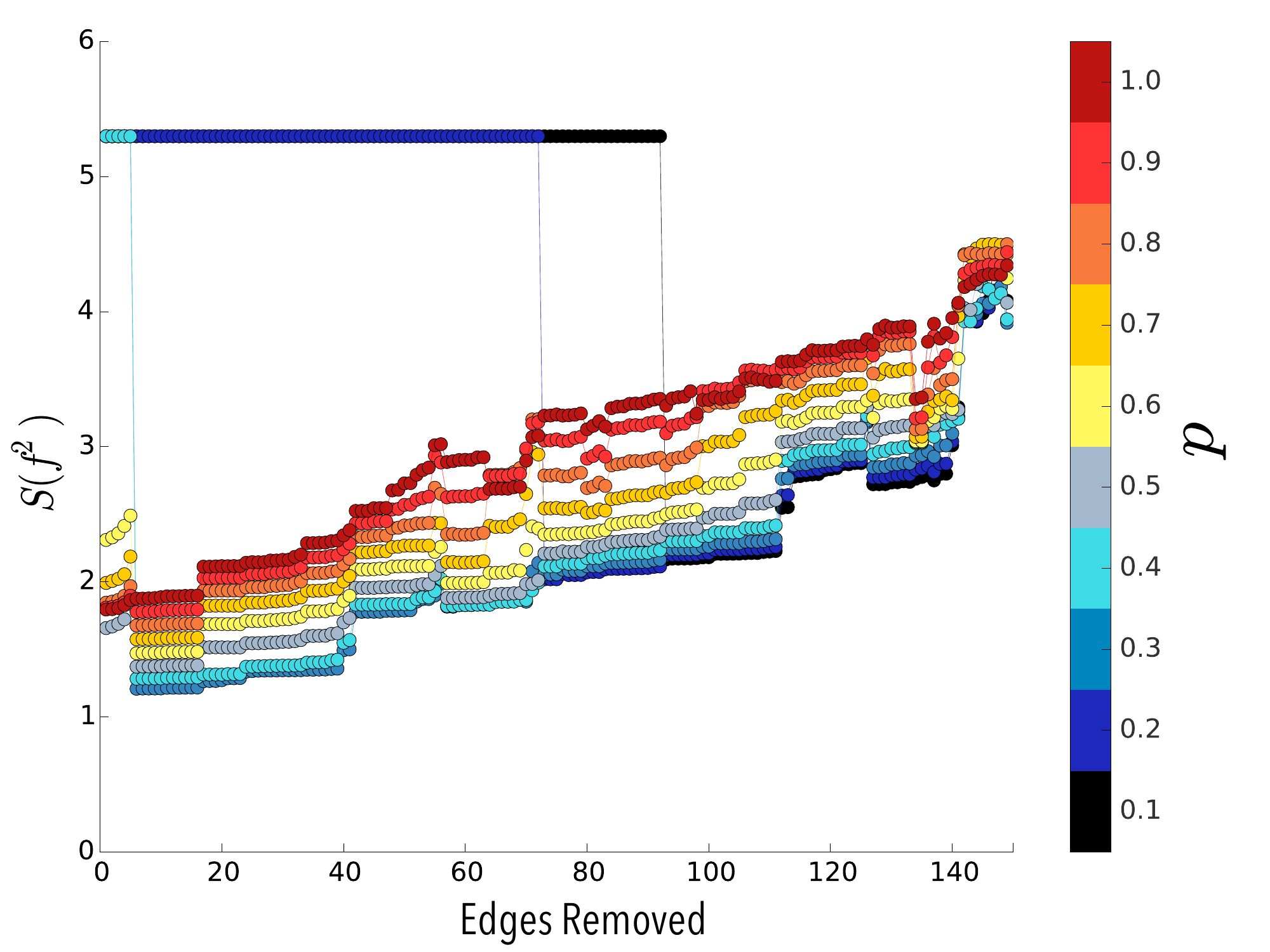}
\includegraphics[width=\columnwidth]{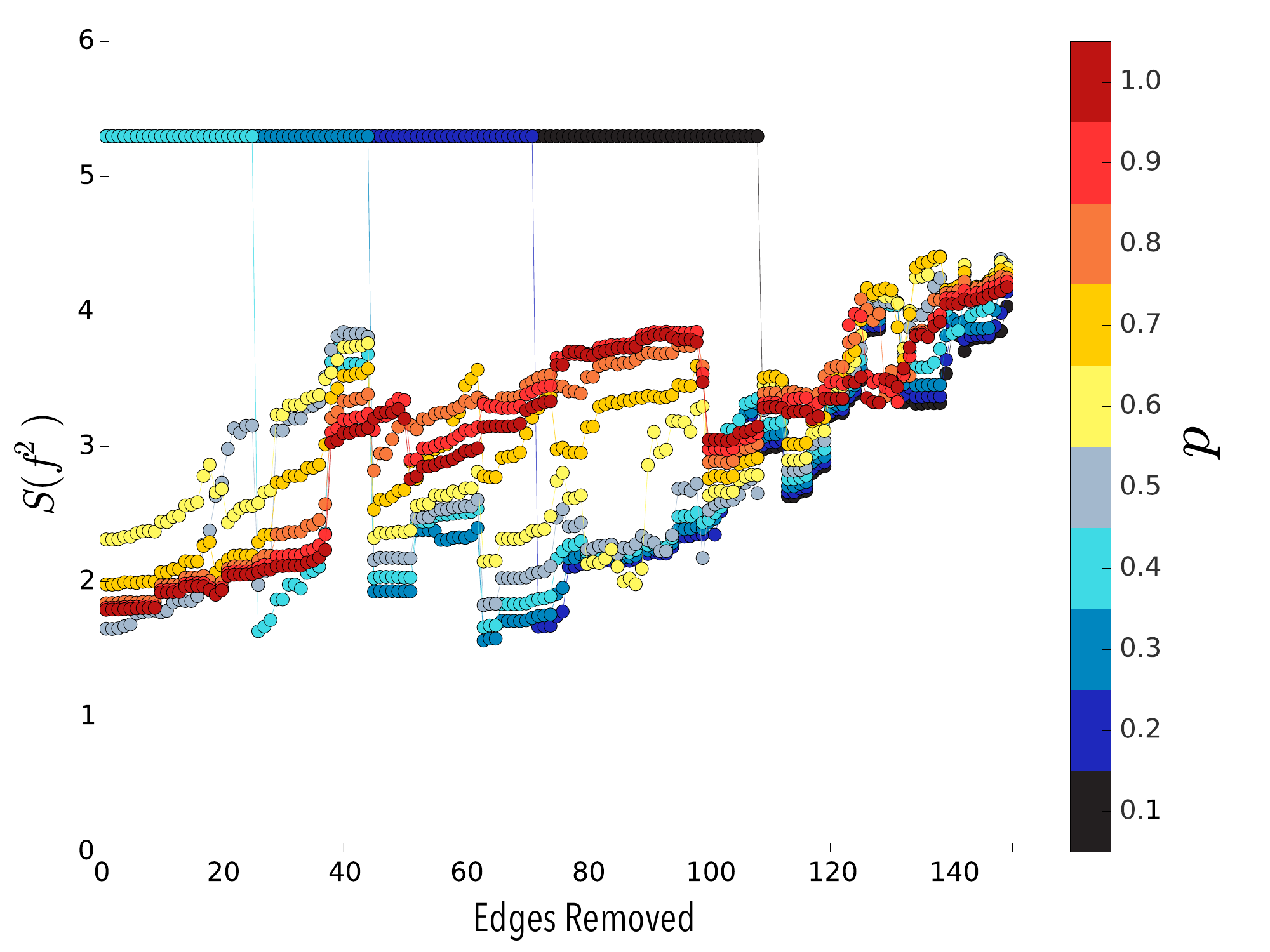}
\caption{Shannon entropy of the Fiedler vector under attack, also called discrete layer degradation, with descending (upper panel) and ascending (lower panel) removals from the ordered links with respect to $r$. The multiplex network used in this experiment is similar to the one used in Fig. \ref{fig:continousdeg}.}
\label{fig:ddfig}
\end{figure} 

In this section, we look to the Fiedler eigenvector, i.e., the eigenvector associated with the algebraic connectivity, in order to understand the mechanism that triggers the transition. Without loss of generality, let us consider a two-layer multiplex network. In the disconnected phase, that is, when $p=p_0$, the Fiedler vector has the form $\bar{x}_2=(1 \dots 1\mid -1,\dots,-1)^T$, i.e., all nodes in the same layer have the same entry of the Fiedler vector. The structural impact on the algebraic connectivity of removing a link in a given layer can be approximated as \cite{Milanese2010}
\begin{equation}
\Delta\bar{\mu}_2=\sum_{ij}(\bar{x}_{2i}-\bar{x}_{2j})=0,
\end{equation}
which is true for $p_0<p^*$. However, the removal of a link has the side effect of lowering $p^*$. Thus, there will be a point in the link removing process at which $p_0$ is no longer lower than the actual $p^*$, the Fiedler vector will have a different structure and thus $\Delta\bar{\mu}\neq 0$, causing a drop in the algebraic connectivity. We remark that these considerations are valid also for the continuous degradation model. These observations motivate us to study the Shannon entropy of the Fiedler vector, which is defined as
\begin{equation}
S=- \sum_i \bar{x}_{2i}^2 \log \bar{x}_{2i}^2, 
\end{equation}
where we consider the Fiedler vector as unitary, i.e., $\| \bar{x}_{2} \| = 1$.
    
As we observe in figures \ref{fig:continousdegentr} and \ref{fig:ddfig}, the entropy of the Fiedler vector starts at its maximum, remains constant and experiments a discontinuous jump that corresponds to the transition. The entropy indicates the level of homogeneity of the  Fiedler vector's components. Note that in the continuous degradation we should analyze Fig. \ref{fig:continousdegentr} from the right to the left. Obviously, it is maximal when all the entries are the same, i.e., before the transition, while after that it reflects the internal organization of the multiplex network. The behavior is identical to the case of the transition in $p$ \cite{DAgostino}, indicating that the Shannon entropy of the Fiedler vector is a good indicator of the structural transition.

\section{Transition-like behavior for no node-aligned multiplex networks}

\begin{figure}[t]
\includegraphics[width=\columnwidth]{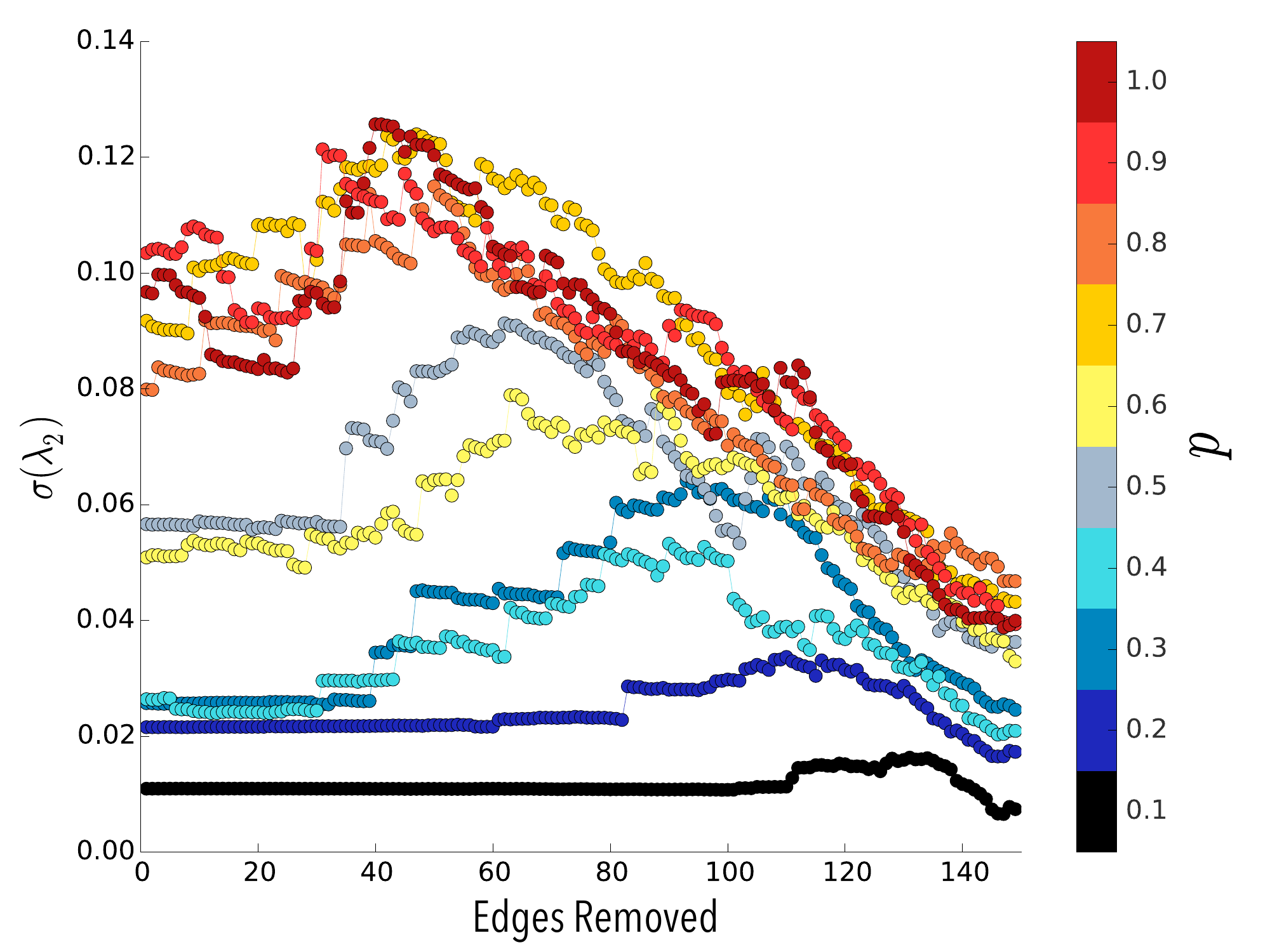}
\caption{Standard deviation of the algebraic connectivity under discrete layer degradation with random removals in a no node-aligned multiplex networks. The multiplex network used in this experiment is similar to the one used in Fig. \ref{fig:continousdeg}.}
\label{rndstdsp}
\end{figure}

\begin{figure}[t]
\includegraphics[width=\columnwidth]{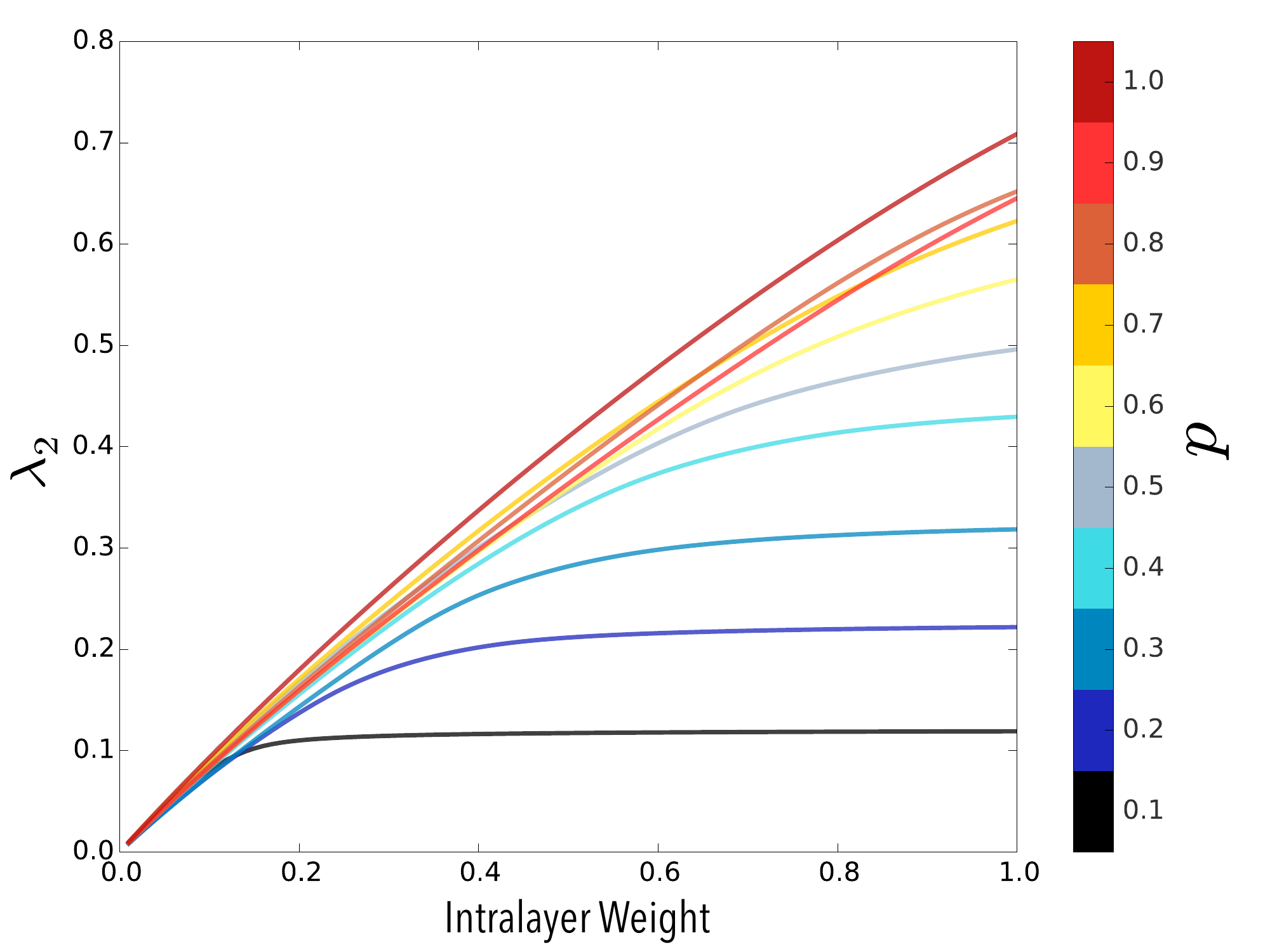}
\includegraphics[width=\columnwidth]{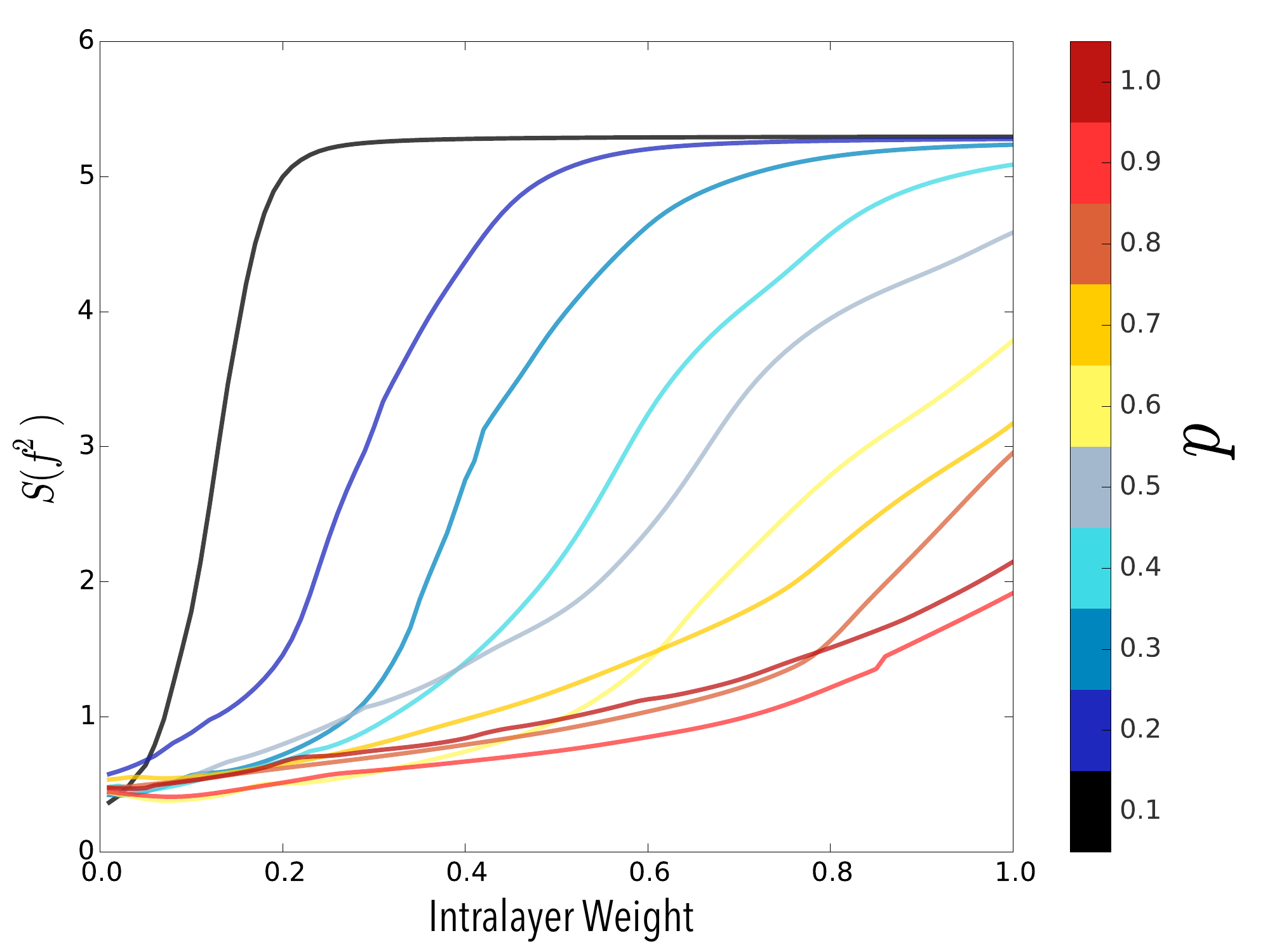}
\caption{Algebraic connectivity of a no node-aligned multiplex (upper panel) and Shannon entropy of the Fiedler vector under contentious layer degradation for a no node-aligned multiplex (lower panel).}
\label{fig:continousdegsparse}
\end{figure}

As previously discussed, the structural transitions, both in $p$ and in $t$, are a consequence of an eigenvalue crossing between the always present eigenvalue $mp_0$ and the next eigenvalue. However, only in the case of node-aligned multiplex networks, $mp_0$ is an eigenvalue of the supra-Laplacian, $\bar{\mathcal{L}}$. On the other hand, for no node-aligned multiplex networks, $mp_0$ is only a bound \cite{Sanchez2014}. Hence, for no node-aligned multiplex networks, a true transition does not exist. Interestingly enough, if we perform a layer degradation by links failure we observe a similar behavior of the standard deviation, see Fig. \ref{rndstdsp}. In the initial regime, the variation is not zero but its fluctuations are approximately constant, while in the final regime the fluctuations follow the same behavior of the true transition of the node-aligned case. 

Furthermore, we observe a phase transition-like behavior also in the case of continuous degradation, as we show in Fig. \ref{fig:continousdegsparse}. In the upper panel, we present the algebraic connectivity, while the lower panel shows the entropy of the Fiedler vector. From the upper panel, two regimes can be roughly recognized, in which the algebraic connectivity is approximately constant and then it varies with a linear trend in $t$. This behavior is a consequence of the fact that $mp$ is only an upper bound for the algebraic connectivity, $\bar{\mu}_2$, as also is $\mu_2(\mathbf{L}_a)$ \cite{Sanchez2014}. In each regime, one bound is sharper than the other. Finally, the same happens to the Shannon entropy of the Fiedler vector, presented in the lower panel of Fig. \ref{fig:continousdegsparse}. Note, however, as expected, that now we cannot observe the jump observed in the node-aligned multiplex case (previously presented in Fig. \ref{fig:continousdeg}).

\section{Conclusions}

In this work we have shown that layer degradation in node-aligned multiplex networks triggers a structural transition of the same kind of the transition induced by varying the coupling parameter $p$. In order to characterize this phenomenon, we studied: (i) failures, modeled as the random removal of edges, (ii) attacks, defined as the removal of edges according to its expected impact on the algebraic connectivity, $r$, and (iii) the continuous layer degradation, where a whole layer weight is lowered, i.e., we decrease all the edge weights associated to that layer. Complementary to the experiments conducted in the analysis and characterization of the algebraic connectivity transition, we also have shown that the Shannon entropy of the Fiedler vector is a suitable measure to quantify the transition point. In addition, we presented evidences of remaining signals of a transition in no node-aligned multiplex networks, where a true transition is not expected.

Finally, note that, despite traditional single-layer networks, in which removing links or lowering their weights will cause in general a finite variation of the algebraic connectivity, for a multiplex network in the disconnected phase the degradation of layers will not affect the algebraic connectivity until a critical point is reached. Importantly, this is not the case for the structural transition triggered by the degradation of the coupling between layers. In this sense, multiplex networks are more resilient to damages and attacks to their layer structure than an isolated layer is if the layers were in the disconnected phase. The results showed here are directly applicable also to regular interdependent networks \cite{ReviewKivela}, since the transition is rooted in an eigenvalue crossing, which also applies to such  cases \cite{Sanchez2014,Cozzo2016Characterization}.

\acknowledgments
GFA acknowledges Fapesp for the sponsorship provided (grants 2012/25219-2 and 2015/07463-1). FAR acknowledges the Leverhulme Trust, CNPq (Grant No. 305940/2010-4) and FAPESP (Grants No. 2016/25682-5 and grants 2013/07375-0) for the financial support given to this research. Y. M. acknowledges support from the Government of Arag\'on, Spain through grant E36-17R (FENOL) and by MINECO and FEDER funds (grant FIS2017-87519-P).

\end{document}